\documentclass[nopacs,twocolumn,prb]{revtex4-1}
\usepackage{amsfonts}
\usepackage{amsmath}
\usepackage{amssymb}
\usepackage{color}
\usepackage{graphicx}
\begin{document}

\title{Majorana fermions in superconducting nanowires without spin-orbit coupling}

\author{Morten Kjaergaard,$^1$, Konrad W\"{o}lms,$^{1,2}$, and Karsten Flensberg$^1$}
\affiliation{$^1$Niels Bohr Institute, University of Copenhagen, Universitetsparken 5, 2100 Copenhagen, Denmark\\
$^2$Fachbereich Physik, Freie Universit\"at Berlin, 14195 Berlin, Germany}

\begin{abstract}
We study nanowires with proximity-induced $s$-wave superconducting pairing in an external magnetic field that rotates along the wire. Such a system is equivalent to nanowires with Rashba-type spin-orbit coupling, with strength proportional to the derivative of the field angle. For realistic parameters, we demonstrate that a set of permanent magnets can bring a nearby nanowire into the topologically non-trivial phase with localized Majorana modes at its ends. This occurs even for a magnetic field configuration with nodes along the wire and alternating sign of the effective Rashba coupling.
\end{abstract}
\date{\today}
\maketitle

Currently, there is an intensive search for materials and geometries that can support topological states, such as Majorana bound states appearing in systems with a $p$-wave superconducting order. One motivation for this effort is the possible applications of Majorana states as the basis for topological quantum computing,\cite{Nayak2008} for example in hybrid structures combined with other qubit systems\cite{Sau2010c,Flensberg2011,Leijnse2011b,Jiang2011,Hassler2011,Bonderson2011} to achieve universal quantum computing. This large activity is inspired by the recent suggestions to use an ordinary $s$-wave superconductor in proximity to a topological insulator\cite{Fu2008} or to systems with strong spin-orbit coupling \textit{and} a strong Zeeman or exchange field to engineer  superconductors with $p$-wave order parameter.\cite{Fujimoto2008,Lee2009,Lutchyn2010,Oreg2010,Alicea2010,Lutchyn2011,Sau2010,Sau2010b,Potter2010}
The role of the spin-orbit interaction is to mix the spin directions, polarized by the applied field, such that there is an overlap with the induced $s$-wave superconducting order. For example, a one-dimensional system with spin-orbit coupling and a Zeeman field perpendicular to the spin-orbit field in proximity to a $s$-wave superconductor\cite{Oreg2010} effectively (when projected to the lowest band) reduces to Kitaev's model,\cite{Kitaev2001} known to support Majorana modes at its ends. The hitherto proposed systems, which combine $s$-wave superconductors, Rashba spin-orbit coupling and splitting of spin degeneracy, include heterostructures of superconductors and ferromagnets, with spin-orbit coupling at the superconductor surface\cite{Lee2009,Potter2010,Sau2010,Duckheim2011} or an additional layer of strong spin-orbit coupling semiconductors with proximity induced superconductivity.\cite{Alicea2010,Sau2010b,Lutchyn2011} A one-dimensional version of this allows the ferromagnet to be replaced by an external magnetic field.\cite{Lutchyn2010,Oreg2010} Other recent one-dimensional suggestions included using nanotubes\cite{Sau2011a} or chains of quantum dots.\cite{Sau2011b}
\begin{figure}[ptb]
\centering
\includegraphics[width=.4\textwidth]{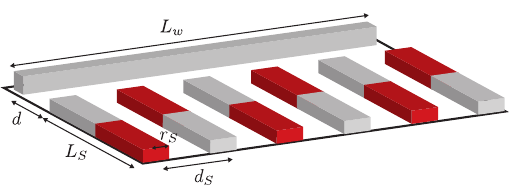}
\caption{ (Color online) Sketch of a quantum wire subjected to a spatially varying magnetic field due to a set of magnetic gates deposited on the substrate.
\label{fig:geo}}
\end{figure}

The realization of Majorana fermions could be easier if one could relax on the requirement of spin-orbit interaction in the material. Braunecker \textit{et al.}\cite{Braunecker2010} showed that in one dimension a spiralling magnetic field is equivalent to a Rashba spin-orbit coupling in the local frame. Such a situation can be realized by a helical nuclear magnetic order.\cite{Braunecker2009} Along the same lines, Choy \textit{et al.}\cite{Choy2011} proposed to use an array of magnetic nanoparticles to create a similar situation, but in this case the spatial dependence of the magnetization direction was due to non-collinear alignment of the magnetic moments of the nanoparticles. Similarly, it was argued that a superconductor with helical magnetism\cite{Martin2011} produces a $p$-wave superconductor.

Here, we investigate a different approach based on spatially varying magnetic fields generated by easily fabricated magnets of submicrometer sizes. To demonstrate the principle, we focus in our calculations on the geometry shown in Fig.~1, namely, a one-dimensional wire next to a set of parallel permanent magnets, with either alternating or parallel magnetizations [see also Fig.~2(a)]. Alternatively, the spatial variations in the Zeeman field direction can be created in a bend wire with an anisotropic $g$-factor and a constant field.

To model such a geometry, let us consider a one dimensional system with a magnetic field that changes direction along the wire. The wire is placed in tunnel contact with a $s$-wave superconductor, which induces a pairing interaction potential $\Delta$ (assumed real and  local below). Using the Nambu four-vector basis $\Psi=(\Psi_\uparrow, \Psi_\downarrow,\Psi_\downarrow^\dagger,-\Psi_\uparrow^\dagger)$, the Hamiltonian for the nanowire is $H=\frac12\int d\xi\Psi^\dagger \mathcal{H}\Psi$, where
\begin{equation}\label{H}
\mathcal{H}=\mathcal{H}_0+\mathcal{H}_S,
\end{equation}
where the normal and superconducting parts are, respectively,
\begin{equation}\label{H0S}
\mathcal{H}_0=\left(\frac{p_{\xi }^{2}}{2m}-\mu\right)\tau_3+\frac{1}{2}g\mu_B^{{}}\mathbf{B}(\xi )\cdot\boldsymbol{\sigma},\quad\mathcal{H}_S=\Delta\tau_1,
\end{equation}
where $\mathbf{B}(\xi )$ is the effective magnetic field field along the wire coordinate $\xi$, $g$ is the effective Land\'{e} $g$-factor, $\sigma _{i}$ ($i=1,2,3$) are Pauli matrices operating in spin-space, while $\tau_i$ are Pauli matrices in electron-hole space. We have not included an intrinsic spin-orbit coupling in the Hamiltonian \eqref{H0S}, but it should be noted that in order to align with the effective spin-orbit field defined below, an intrinsic spin-orbit field should be perpendicular to the plane of the applied field.

Below, we present numerical calculations of the Hamiltonian in Eq.~(\ref{H0S}) for a magnetic field configuration set up by a set of permanent magnets, causing a spiralling direction of the field. To understand why such a system is equivalence to a Rashba-type spin orbit interaction, we follow Braunecker \textit{et al.}\cite{Braunecker2010} and perform a unitary transformation. We rotate the $z$-axis of the local spin basis to the direction of the magnetic field by the unitary operator $U=\exp(i(\varphi/2)\sigma_{xy})$, where $\sigma_{xy}=((\mathbf{B}\times \hat{\mathbf{z}})\cdot\boldsymbol{\sigma})/|\mathbf{B}\times \hat{\mathbf{z}}|$ and $\cos\varphi=\mathbf{B}\cdot\hat{\mathbf{z}}/B$, with $B=|\mathbf{B}|$. The transformed Hamiltonian $\tilde{\mathcal{H}}=U^{\dagger }\mathcal{H}U$ then becomes
\begin{equation}\label{tildeH}
\tilde{\mathcal{H}}_0=\left(\frac{1}{2m}p_{\xi }^{2}-\mu\right)\tau_3+\tilde{\mathcal{H}}_R+\tilde{\mathcal{H}}_2+\frac{g\mu_B^{{}}B}{2}\sigma _{3},
\end{equation}
while the rotation, of course, leaves the singlet pairing interaction invariant $\tilde{\mathcal{H}}_S=\mathcal{H}_S$. The rotation generates two new terms, $\tilde{\mathcal{H}}_R=(\hbar/mi)U^\dagger U'p_\xi\tau_3$ and $\tilde{\mathcal{H}}_2=(-\hbar^2/2m)U^\dagger U''\tau_3$, where the primes denote differentiation with respect to $\xi$. The term $\tilde{\mathcal{H}}_R$, being proportional to the momentum operator, is a Rashba-type spin-orbit coupling that reduces to
\begin{equation}\label{Hrashba}
\tilde{\mathcal{H}}_R=\frac{\hbar}{m}
    \left(\frac12 \sigma_{xy}\frac{d\varphi}{d\xi}+U^\dagger\frac{d\sigma_{xy}}{d\xi}\sin(\varphi/2)\right)p_\xi\tau_3.
\end{equation}
This simplifies further  if the field lines and the wire lie in a single plane, which is indeed the case for the lateral structures considered here (see Figs.~1 and 2). In that situation, we can choose $\hat{\mathbf{z}}$ to be in the plane of the field and hence $\sigma_{xy}'=0$. To avoid a sign change of $\sigma_{xy}$ when $\mathbf{B}\parallel\hat{\mathbf{z}}$, we choose  $U=\exp(i\varphi \sigma_\perp/2)$, with $\sigma_\perp$ constant and $\varphi$ continuous. For the lateral configuration, the Rasbha-type spin-orbit coupling part of the Hamiltonian thus becomes
\begin{equation}\label{tildeHR}
\tilde{\mathcal{H}}_R=\alpha_\mathrm{eff}\sigma_\perp p_\xi\tau_3,
\end{equation}
where we define an effective spin-orbit interaction coefficient as
\begin{equation}\label{aeff}
\alpha_{\mathrm{eff}} =\frac{\hbar}{2m}\frac{d\varphi}{d\xi}.
\end{equation}
The second new term in the transformed Hamiltonian \eqref{tildeH} becomes $\tilde{\mathcal{H}}_2=(\hbar^2/2m)[(\varphi'/2)^2-i\varphi''\sigma_\perp/2]\tau_3$, where the first term renormalizes the chemical potential. The last term has the form of an imaginary magnetic field parallel to the plane of the spin-orbit field, and it appears because the total Hamiltonian has to be Hermitian. Alternatively, this term can be absorbed into  Eq.~\eqref{tildeHR} by writing it in symmetrized form, \textit{i.e.}, $(\alpha_\mathrm{eff}p_\xi+p_\xi\alpha_\mathrm{eff})/2$. In order to drive the one-dimensional superconductor into the $p$-wave state, a large spin-orbit coupling and a Zeeman field perpendicular to the spin-orbit field are desirable.\cite{Oreg2010} Therefore the rotation of the magnetic field should be optimized to have a large first derivative, $\varphi'$.
\begin{figure}[ptb]
\centering
\includegraphics[width=.5\textwidth]{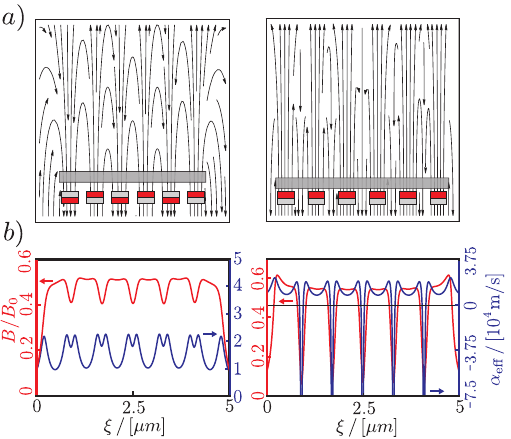}
\caption{ (Color online) Sketch of two realistic systems having a series of permanent magnets with either alternating (left) or parallel (right) magnetization directions. The top panels show the field lines and it is clearly seen that the field changes direction a number of times along the wire. The bottom panels show the amplitude of the magnetic field $B/B_0$ in red (light gray), as well as the effective induced spin-orbit coupling $\alpha_\mathrm{eff}$ in blue (dark gray) (calculated with effective mass $m=0.014m_e$). The overall scale for the magnetic field $B_0$ is set by the magnetization of the magnets. The magnets have widths 600 nm, heights 330 nm, and the gap between them is 200 nm. For the left configuration the distance to the wire is 100 nm, and 50 nm for the parallel configuration.
\label{fig:alphaeff}
\label{fig:fieldlines}}
\end{figure}

An optimal and illustrative example is therefore a sinusoidally rotating magnetic field $\mathbf{B}(\xi )=B_c(\sin(\xi/R) ),0,\cos(\xi/R))$, which gives the transformed Hamiltonian\cite{Braunecker2010}
\begin{equation}\label{Hsinus}
\tilde{\mathcal{H}}_{0}^{\mathrm{sine}}= \left(\frac{p_{\xi }^{2}}{2m}-\mu+\frac{\hbar }{2mR}\sigma
_{2}p_{\xi }^{{}}+\frac{\hbar ^{2}}{8mR^{2}}\right)\tau_3+\frac{g}{2}\mu_B^{{}}B_c\sigma
_{3}.
\end{equation}
This model is seen to be identical to that of a one-dimensional wire with Rashba spin-orbit coupling, with $\alpha_{\mathrm{eff}}=\hbar/2mR$. To have a large effective spin-orbit coupling, one should therefore use low mass materials and engineer a large curvature for the magnetic field. Using parameters relevant for InSb (which also has a large $g$-factor, $g\approx 50$), we set $m=0.014 m_{e}$, which together with $R=100$nm gives $\alpha_\mathrm{eff}\approx 3\times 10^{4}$ m/s. This is similar to the spin-orbit coupling strength in, for example, InAs.

A realistic way to generate a magnetic field that rotates along the wire is to place the wire next to a set of permanent magnets, as illustrated in Fig.~1. The magnets could, for example, be made of Co. The left configuration in Fig.~2(a) has a staggered configuration (which could be realized by the magnets having different sizes, so that their hysteresis loops are also different), while the configuration to the right has aligned magnets. We have computed the fields, with dimensions that are easily fabricated with micro technology. For both cases, Fig.~2(b) shows the amplitude of the magnetic fields $B$ (in units of $B_0$, set by the permanent magnets) and the effective Rashba spin-orbit coupling parameters $\alpha_\mathrm{eff}$ given by Eq.~\eqref{aeff}. The configuration with alternating magnetization directions is seen to give a field close to that of the optimal sinusoidal variation (which has constant $B$ and $\alpha_\mathrm{eff}$). For the configuration with parallel magnetizations, the effective Rashba coupling changes sign along the wire, but nevertheless with a sufficient integrated weight to induce a phase transition to a state with Majorana fermions, as shown below.

We now discuss the conditions for bringing the nanowire with proximity induced pairing interaction into the topologically non-trivial regime, and thus to have Majorana fermions at its ends. Starting with the case of a superconducting wire with sinusoidally rotating magnetic field, leading to the normal part in Eq.~\eqref{Hsinus}, there is a transition to a non-trivial phase when\cite{Oreg2010}
\begin{equation}\label{trans}
    g\mu_B|B_c|>\sqrt{|\Delta|^2+(\mu-\hbar^2/8mR^2)^2},
\end{equation}
where $\mu$ is the chemical potential. It is interesting to note that the curvature term shifts the chemical potential, allowing for a slightly larger density of electrons in the non-trivial phase.

In the general case, the transition to the non-trivial phase cannot be found analytically and we instead resort to numerical methods; guided by the above conclusions, that it is desirable to have a fast rotation of the direction of the magnetic field. In our numerical study we use original Hamiltonian in Eqs.~\eqref{H} and \eqref{H0S}.
\begin{figure}[ptb]
\centering
\includegraphics[width=.4\textwidth]{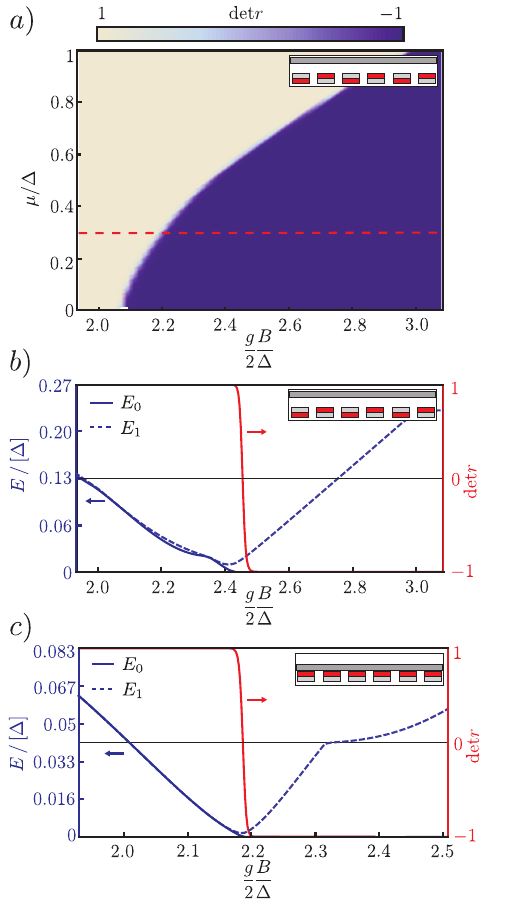}
\caption{ (Color online) Phase diagram for the superconducting nanowire subject to the magnetic field in the two configurations shown in Fig.~\ref{fig:fieldlines}, calculated for $5-\mu$m-long wire with $m=0.014 m_e$ and $\Delta=$ 0.3meV. The topologically non-trivial phase occurs when the determinant of the reflection changes from 1 to $-1$, which is shown in the phase diagram in (a) for changing magnetic field $B$ and chemical potential $\mu$ and also in the lower panels along the dashed (red) line $\mu=0.3\Delta$. The lower panels show the lowest positive eigenvalue $E_0$, which goes to zero after the topological phase transition, as well as the next positive eigenvalue $E_1$. After the transition the difference between the two is equal to the gap of the topologically non-trivial state. Both configurations in Fig.~2 are seen to exhibit a phase transition. In (b) and (c) the lines in red (light gray) show the value of the determinant of the reflection matrix.
\label{fig:phasediagram}}
\end{figure}
We have studied two criteria for the existence of the end modes, namely, that (i) a change of the determinant of the zero-energy  reflection matrix from 1 to $-1$ at the transition to the topological non-trivial phase,\cite{Akhmerov2011,Choy2011} and (ii) the existence of a pair of zero modes. We calculate the scattering matrix of the wire connected to two single-mode metallic leads using the expression\cite{Aleiner2002}
\begin{equation}\label{S}
    S(0)=S_0^{{}}\frac{1+i\pi\nu W^\dagger \mathcal{H}^{-1} W}{1-i\pi\nu W^\dagger \mathcal{H}^{-1}W}S_0^T,
\end{equation}
where $W$ is the coupling matrix between to the two leads (with density of states $\nu$) at the end point of a discretized version of the Hamiltonian and it is thus a matrix with dimension $2\times N$, with $N$ being the number of sites in the chain, having entries only at the upper left and lower right corners. Further, $S_0S_0^T$ is the $S$-matrix in the absence of coupling to the wire. For the numerical calculations, the wire is discretized as a tight-binding chain with hopping matrix element $t$ between sites separated by $a$ and the chemical potential in the bottom of the band with approximate quadratic dispersion, so that we can relate to the effective mass via $m=\hbar^2/2ta$. Because of finite-size effects the result for the reflection matrix determinant is not independent of the choice of $\nu$, and for the actual calculation we use a value showing most clearly the transition. In Fig.~3 we use $N=200$ and have checked that the results do not change significantly by changing $N$.

In addition, we have numerically diagonalized the discretized version of the Hamiltonian in Eq.~\eqref{H0S} and the topological phase transition is characterized by the point where the system supports a zero-energy state, while having a gap for the continuum states. Figure \ref{fig:phasediagram} shows the results for determinant of the reflection matrix together with plots of the lowest eigenenergies and the gap. In Fig.~\ref{fig:phasediagram}(a) the determinant is seen to have a sharp transition from 1 to $-1$. At the same point where this transition occurs, a zero mode appears. The zero mode continues into the non-trivial phase, while the gap (shown as the energy of the next lowest positive eigenenergy) increases for larger values.

Interestingly, the configuration with parallel magnets also shows a topological phase transition, with similar values of the gap and the necessary magnetic field, see Fig.~3(c) (it should be noticed, however, that for this case the wire should be closer to the magnets to compensate for the partial cancellation of the field lines coming from neighboring magnets). This is interesting because it shows that Majorana fermions can be generated even if the magnitude of the magnetic $B$ goes below the critical field in a small region of space and even if the effective Rashba spin-orbit has an opposite sign in this region--see the lower right panel of Fig.~2 for the detailed shapes. How robust the topological phase is to having regions  with subcritical magnetic field is an interesting question that deserves further studies.
\begin{figure}[ptb]
\centering
\includegraphics[width=.5\textwidth]{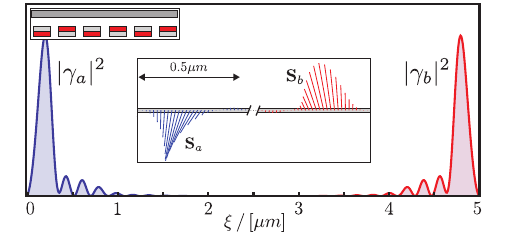}
\caption{(Color online) The localized Majorana modes and their spin texture. The top panel shows the total weight of the end Majorana modes, $\gamma_a$ and $\gamma_b$, while the lower panel depicts the local spin direction for the configuration of alternating magnetizations (Fig.~2 left). The length of the arrows in the inset corresponds to the amplitude of the wave function. Notice that the $x$-axis of the inset is rescaled.
\label{fig:pindsvin}}
\end{figure}

Finally, we investigate the structure of the Majorana fermions. Each end of the wire supports one Majorana state, which we denote $\gamma_a$ and $\gamma_b$. In the four-spinor Nambu basis $\gamma=(u_\uparrow, u_\downarrow,v_\downarrow,-v_\uparrow)$, where $u$ and $v$ are the electron and hole amplitudes, respectively, a Majorana state has $u_\uparrow=v_\downarrow^*$ and $u_\downarrow=v_\uparrow^*$.\cite{selfadjoint} However, for our finite system the Majorana states are coupled by an overlap exponentially small in the distance between them.  Therefore the numerically determined eigenstates are linear superpositions of $\gamma_a$ and $\gamma_b$ with a small finite energy. If the numerical eigenstates are denoted $\gamma_1$ and $\gamma_2$, we generate the states that would correspond to Majorana states for an infinitely long wire as $\gamma_{a}=(\gamma_1+\gamma_2)/\sqrt2$ and $\gamma_{b}=i(\gamma_1-\gamma_2)/\sqrt2$. These states are plotted in Fig.~\ref{fig:pindsvin}. The top panel shows the total weight $|\gamma_{a/b}|^2$ of the two Majorana states localized at the ends of the wire.
It is also interesting to look at the spin direction of the localized Majorana modes,\cite{Bena2011} which is also relevant for a recent proposal of transfer of quantum information between Majorana qubits and spin qubits.\cite{Leijnse2011b} In Fig.~\ref{fig:pindsvin}, we plot the local spin direction of the electron part defined as (see Ref. \onlinecite{Bena2011}) $   \mathbf{S}_{a/b}(\xi)=\langle \gamma_{a/b}|\xi\rangle\langle\xi|\boldsymbol\sigma \otimes \frac12 (1+\tau_z)|\gamma_{a/b}\rangle.$  If one compares the spin direction with the field lines in Fig.~\ref{fig:fieldlines}, it can be seen that the direction is dictated by the direction of the magnetic field and can therefore be tuned. The spin polarization of the Majorana modes results in spin-specific tunnel coupling, which can be used for both manipulation as in Refs.~\onlinecite{Flensberg2011,Leijnse2011b} and detection as in Ref.~\onlinecite{Bena2011}.

In summary, we have shown that quantum wires with proximity induced superconductivity and a spatially varying magnetic field created by a realistic configuration of permanent magnets can have topological excitations in the form of Majorana fermions. This system has advantages over proposals that require a specific form of Rashba spin-orbit coupling. Moreover, by choosing materials with a large $g$--factor, a modest field is sufficient to meet the condition that the Zeeman exceeds the induced pairing potential. Finally, we studied the Majorana states spin texture, which can be tuned by the magnetic field configuration.

\acknowledgments We thank R. Egger, C. Marcus, and A. Reynoso for valuable discussions. The work was supported by The Danish Council for Independent Research $|$ Natural Sciences.


\begin{thebibliography}{10}
\bibitem{selfadjoint}
A Majorana fermion is in second quantization represented by a self-adjoined
  operator $\Psi^\dagger=\Psi$. In our first quantization notation here it
  translates to $\gamma=\sigma_2\tau_2\gamma^*$.
\end{thebibliography}

\end{document}